\documentclass[reprint,twocolumn,secnumarabic,amssymb, nobibnotes, aps, prl,superscriptaddress]{revtex4-1}

\usepackage{extarrows}
\usepackage{amsmath}    % need for subequations
\usepackage{graphicx}    % need for figures
\usepackage{verbatim}    % useful for program listings
\usepackage{color}           % use if color is used in text
\usepackage{subfigure}   % use for side-by-side figures
\usepackage{hyperref}    % use for hypertext links, including those to external documents 
\usepackage{verbatim}  % and URLs
\begin{document}

\title{ Continuum Charge Excitations in High-Valence Transition-Metal Oxides \\ Revealed by Resonant Inelastic X-ray Scattering }
\author{Atsushi Hariki}
%\thanks{hariki@ifp.tuwien.ac.at}
\affiliation{Institute for Solid State Physics, TU Wien, 1040 Vienna, Austria}
\author{Mathias Winder}
\affiliation{Institute for Solid State Physics, TU Wien, 1040 Vienna, Austria}
\author{Jan Kune\v{s} }
%\thanks{kunes@ifp.tuwien.ac.at}
\affiliation{Institute for Solid State Physics, TU Wien, 1040 Vienna, Austria}
\affiliation{Institute of Physics,Czech Academy of Sciences, Na Slovance 2, 182 21 Praha 8, Czechia}

\date{\today}%a

\begin{abstract}
We present a theoretical investigation of the origin of Raman-like and fluorescencelike (FL)
features of resonant inelastic x-ray scattering (RIXS) spectra. Using a combination of 
local-density approximation + dynamical mean-field theory and a configuration
interaction solver for Anderson impurity model, we calculate the $L$-edge RIXS and
x-ray absorption spectra of high-valence transition-metal oxides LaCuO$_3$ and NaCuO$_2$.
We analyze in detail the behavior of the FL feature and show
how it is connected to the details of electronic and crystal structure. On the
studied compounds we demonstrate how material details 
determine whether the electron-hole continuum can be excited in the $L$-edge RIXS process.
\end{abstract}

\maketitle 
%%%%%%%%%%%%%%%%%%%%%%%%%%%%%%%%%%%%%%%%%%%%%%%%%%%%%%%%%%%%%%%%%%%
%%%%%%%%%%%%%%%%%%%%%%%%%%%%%%%%%%%%%%%%%%%%%%%%%%%%%%%%%%%%%%%%%%%

The excitation spectrum is a fundamental characteristic that determines 
properties of a physical system. While the excitations of weakly correlated electrons can be
built out of elementary ones and therefore different spectroscopic experiments show the similar spectra, e.g., the same gap for charge, spin and optical excitations, correlated materials
such as transition metal oxides (TMO) are different~\cite{imada98,khomskii14}. 
Here, the connection between various types of excitations is buried deep in the experimentally unobservable wave function and different techniques are necessary to probe specific excitations.

Following the remarkable improvements of energy resolution in the past decade
resonant inelastic x-ray scattering (RIXS) became a popular tool to study diverse materials~\cite{Ament11}. Its sensitivity to a range of two-particle excitations
enables observation of low-energy spin, orbital and charge excitations ($\sim$100 meV)~\cite{Betto17,Fabbris17,Kim12,Braicovich09} together with
high-energy excitations ($\sim$1$-$10 eV), such as atomic multiplets 
or charge-transfer (CT) excitations~\cite{Ghiringhelli05,matsubara05,groot_kotani}.
The versatility of RIXS is paid for complicated interpretation of the resonant spectra
which requires theoretical modeling.

Two features have been observed in the 
$L$-edge RIXS ($2p$$\to$$3d$$\to$$2p$) of TMO when scanning the incident 
photon energy $\omega_{\rm in}$ across the x-ray absorption spectra (XAS):
a Raman-like (RL) signal with a constant energy loss 
$\omega_{\rm loss}=\omega_{\rm in}-\omega_{\rm out}$ and 
a fluorescencelike (FL) signal with a constant emission energy 
$\omega_{\rm out}$, i.e.,  a linear dependence of  $\omega_{\rm loss}$ on $\omega_{\rm in}$~\cite{Ament11,Kotani01,matsubara05,Zhou11,Schmitt04,Pfaff18}.
The analysis of material specific behavior of these features shows a potential 
for addressing questions concerning itinerancy  of charge carriers and localization
of charge excitations~\cite{Zhou11,Bisogni16,Pfaff18}. While several experimentally 
motivated interpretations were put forward,
unified description of the atomic-like RL and itinerant FL 
features poses a theoretical challenge~\cite{Vinson2011,Gilmore2016}.

The $L$-edge RIXS spectrum of the CT Mott insulator NiO~\cite{Ghiringhelli05,matsubara05} 
exhibits a RL behavior at $\omega_{\rm in}$ of the main absorption peak well separated from the FL feature at $\omega_{\rm in}$ of the CT satellite. A different behavior was reported in the negative-CT compound NdNiO$_3$, where Bisogni {\it et al.}~\cite{Bisogni16} observed
merging of the low-$\omega_{\rm loss}$ RL and FL features in Ni $L_3$-RIXS.
Moreover, the details of low-$\omega_{\rm loss}$ FL feature exhibit distinct temperature dependence connected to opening of charge gap at low temperatures.
Bisogni {\it et al.} interpreted 
the low-$\omega_{\rm loss}$ FL feature as a signature of unbound particle-hole pairs in
the RIXS final state.
Zhou~{\it et al.}~\cite{Zhou11} studied $L$-edge RIXS in LaAlO$_3$/SrTiO$_3$ heterostructures and interpreted the relative intensity of the FL feature as a measure of
itinerant carrier concentration. This conclusion was recently questioned by Pfaff $et$ $al$.~\cite{Pfaff18} who suggested that either RL or FL signals reflect the nature of the intermediate state of the RIXS process.

%Theoretical description of RIXS in high-valence TMO is a challenging task.
The commonly used cluster model with the TM 3$d$ and the neighboring O 2$p$ orbitals
misses the FL feature completely due the lack of continuum (delocalized) particle-hole excitations.
This is remedied by the Anderson impurity model (AIM);~however, the use
of simple ad hoc bath densities of states~\cite{Ghiringhelli05,matsubara05,groot_kotani,Bisogni16}
does not allow to capture the important material details.
To overcome this limitation we employ the
local-density approximation (LDA) + dynamical mean-field theory (DMFT)~\cite{kotliar06,georges96}
approach.
%The LDA+DMFT calculation brings an accurate hybridization function of AIM, in which the many-body valence states are embedded. 
The AIM with material specific hybridization density is then extended to include the core orbitals~\cite{Hariki17}.
%By construction, besides the atomic multiplet excitations,
%the charge excitations, coupled to the TM 3$d$ states via the hybridization, are described by the AIM accurately.
%Thus an accurate description of the CT excitation, triggered by the X-rays, is achieved. 
This approach, which we recently applied to study nonlocal screening effects
in $L$-edge x-ray photoemission spectroscopy (XPS)~\cite{Hariki17}, is here extended to analyze RIXS.
%by considering the coherent second-order optical process properly.

\begin{figure}
%\vspace{-1.0cm}
\begin{center}
    \includegraphics[width=75mm]{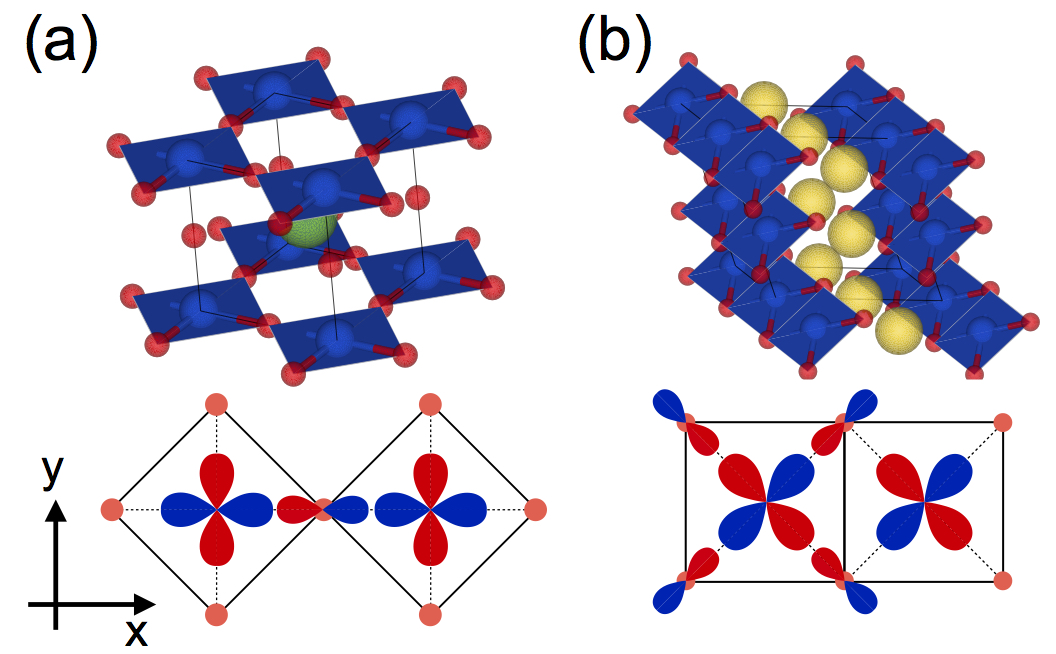}
\end{center}
\vspace{-0.4cm}
\caption{The crystal structures of (a) LaCuO$_3$ with $P4/m$ and and $C2/m$ space groups in paramagnetic
and antiferromagnetic states, respectively~\cite{Bringley93} 
(b) NaCuO$_2$~\cite{Nathaniel89} visualized by VESTA~\cite{vesta}.
The blue, red, green, and yellow circles represent Cu, O, La and Na atoms, respectively.
The sketch of the $xy$ plane is shown together.}
\label{fig_crys}
\end{figure}

In order to focus on the physics of RIXS excitations, we
avoid the uncertainties brought about by structural phase transition in nickelates
and choose two isoelectronic Cu$^{3+}$ compounds LaCuO$_3$ and NaCuO$_2$
as model systems for high-valence TMO. Both share a tiny CT energy~\cite{Mizokawa91,Khomskii01,Mizokawa98,Czyifmmod94,Mizokawa94}
leading to small or no gap~\cite{Mizokawa91,Mizokawa98,Czyifmmod94,Bringley93,Darracq93}.
%e.g.~small gap insulator with $p$-type character. % and very small CT energy.
As we show later the key difference between the two compounds is the lattice geometry
with corner-sharing CuO$_6$ octahedra in LaCuO$_3$, but edge-sharing ones in 
NaCuO$_2$;~see Fig.~\ref{fig_crys}.

%%%%%%%%%%%%%%%%%%%%%%%%%%%%%%%%%%%%%%%%%%
%%%%%%%%%%%%%%%%%%%%%%%%%%%%%%%%%%%%%%%%%%

%%%%%%%%%%%%%%%%%%%%%%%%%%%%%%%%%%%%%%%%%%%%%%%%%%%%%%%%%%%%%%%%%%%
%%%%%%%%%%%%%%%%%%%%%%%%%%%%%%%%%%%%%%%%%%%%%%%%%%%%%%%%%%%%%%%%%%%

The calculation proceeds in two steps. First, a standard LDA+DMFT calculation is performed
as follows. The LDA band structure obtained with Wien2K~\cite{wien2k}
%for $P4/m$ tetragonal and $C2/m$ monoclinic structures for LaCuO$_3$~\cite{Bringley93} and NaCuO$_2$~\cite{Nathaniel89}, respectively. 
is projected~\cite{wien2wannier,wannier90} onto a $dp$ tight-binding model spanning the Cu 3$d$ and O 2$p$ orbitals and augmented
with the electron-electron interaction within the Cu 3$d$ shell, giving the Hamiltonian
%%%%%%%%%%%%%%%%%%%%%%%%%%%%%%%%%%%%%%%%%%%%%%%%
\begin{equation}
 H=\sum_{\textit{\textbf{k}}} 
  \begin{pmatrix}
      \textit{\textbf{d}}^{\dag}_\textit{\textbf{k}} & \textit{\textbf{p}}^{\dag}_\textit{\textbf{k}}
 \end{pmatrix}
 \begin{pmatrix}
     h^{dd}_{\textit{\textbf{k}}}-\mu_{\rm dc} & h^{dp}_{\textit{\textbf{k}}} \\
     h^{pd}_{\textit{\textbf{k}}} & h^{pp}_{\textit{\textbf{k}}}
 \end{pmatrix}
   \begin{pmatrix}
      \textit{\textbf{d}}_\textit{\textbf{k}} \\
      \textit{\textbf{p}}_\textit{\textbf{k}}
 \end{pmatrix}
 +\sum_{i}W^{dd}_{i}.
\notag
\end{equation}
%%%%%%%%%%%%%%%%%%%%%%%%%%%%%%%%%%%%%%%%%%%%%%%%
Here, $\textit{\textbf{d}}^{}_\textit{\textbf{k}}$ ($\textit{\textbf{p}}^{}_\textit{\textbf{k}}$)
is an operator-valued vector whose elements
are Fourier transforms of $d_{\alpha i}$  ($p_{\gamma i}$), that annihilate the
Cu 3$d$ (O 2$p$) electron in the orbital $\alpha$  ($\gamma$) in the $i$ th unit cell.
The on-site Coulomb interaction $W_{i}^{dd}$ is parametrized in the usual way~\cite{pavarini1,pavarini2}
with $U=7.5$ and $J=0.98$~eV, typical for Cu systems~\cite{anisimov91}.
%is approximated by the density-density form 
%with parameters $U=7.5$~eV and $J=0.98$~eV,
%which are typical values for Cu systems~\cite{anisimov91}. %and gives 
%The configuration-averaged Coulomb interaction is given by $U_{dd}$=6.9~eV~[ref].
The double-counting term $\mu_{\rm dc}$, which corrects for the $d$--$d$ interaction
present in the LDA step, renormalizes the $p$--$d$ splitting and thus the CT energy.
While several {\it ad hoc} schemes exist to compute $\mu_{\rm dc}$ (with somewhat different results), we treated $\mu_{\rm dc}$ as adjustable parameter fixed by comparison to the available $L$-edge XAS and valence XPS data, see the Supplementary Material (SM)~\cite{sm}. The strong-coupling continuous-time quantum Monte Carlo method~\cite{werner06,boehnke11,hafermann12} with density-density approximation to the on-site interaction was used to solve the auxiliary AIM.

%The relation of $\mu_{\rm dc}$ and the CT energy is also explained in SM.
%As an impurity solver in the DMFT self-consistent loop, 
%we use the continuous-time quantum Monte Carlo method in the hybridization expansion algorithm
%with improved estimator techniques~\cite{werner06,boehnke11,hafermann12}.
%For LaCuO$_3$, we also study the $G$-type antiferromagnetic (AF) ordering
%obtained by allowing the spin dependence of the self-energy in the magnetic unit cell.

%%%%%%%%%%%%%%%%%%%%%%%%%%%%%%%%%%%%%%%%%%
\begin{figure}
%\vspace{-1.0cm}
\begin{center}
    \includegraphics[width=87mm]{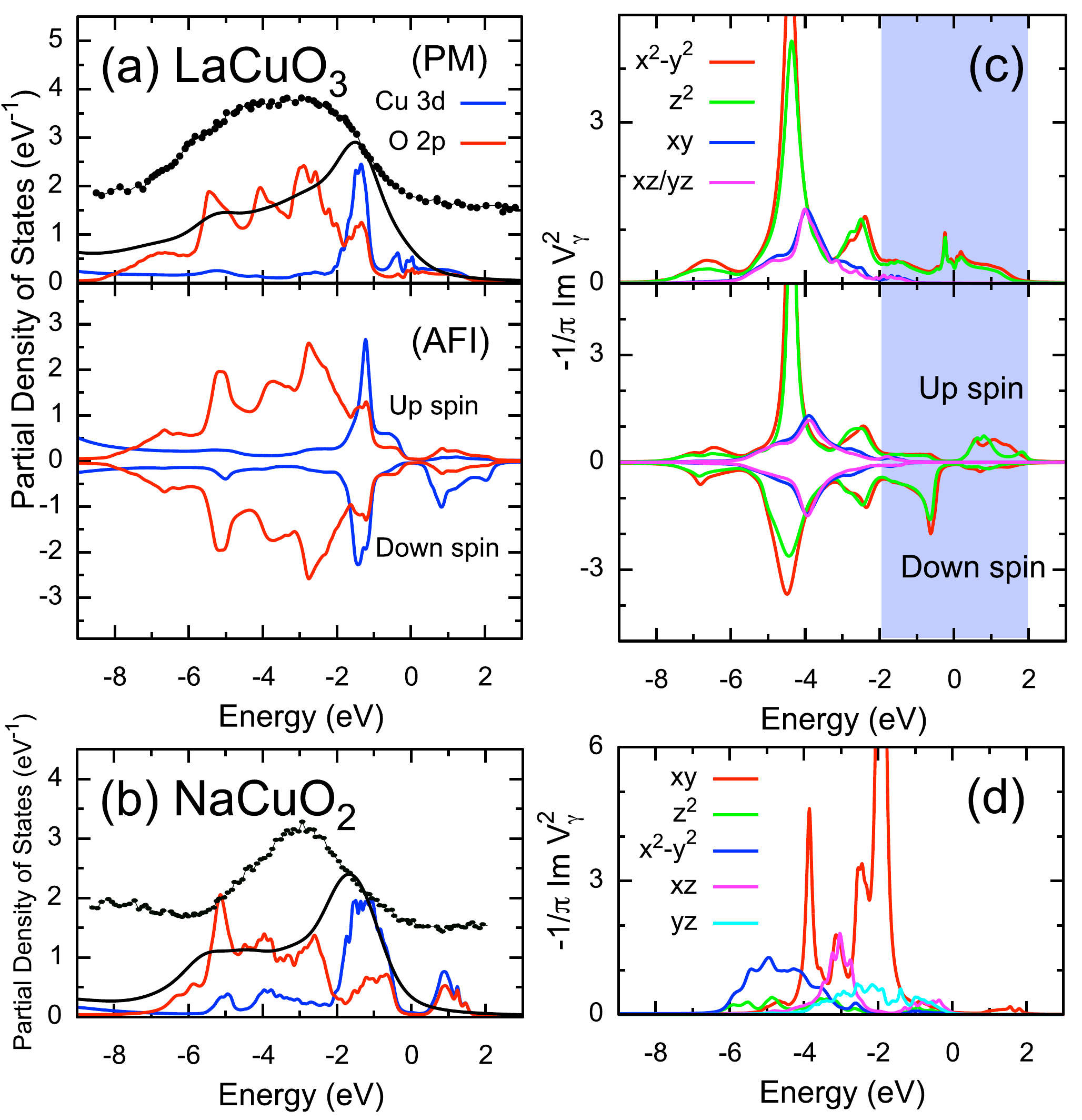}
\end{center}
\vspace{-0.4cm}
\caption{\label{fig_hyb}LDA+DMFT 1P density of states for (a) LaCuO$_3$ in the PM and AFI phases and (b) NaCuO$_2$.
The hybridization function $V_{\gamma}^2(\varepsilon)$ of LaCuO$_3$ (c) in the PM and AFI phases and NaCuO$_2$ (d).
The energy origin is taken at $E_F$. 
%The experimental XPS data \textcolor{red}{(symbols)} of LaCuO$_3$ and NaCuO$_2$, taken from Refs~\cite{Mizokawa98} and \cite{Mizokawa94}, respectively, \textcolor{red}{are compared to the theoretical ones (black line)
The experimental XPS data (symbols) of LaCuO$_3$~\cite{Mizokawa98} and NaCuO$_2$~\cite{Mizokawa94}
are compared to the theoretical ones (black line)~\cite{cross}.
$\mu_{\rm dc}$=55.64~eV is employed in the calculation.
}
\vspace{-0.25cm}
\end{figure}
%%%%%%%%%%%%%%%%%%%%%%%%%%%%%%%%%%%%%%%%%%

%%%%%%%%%%%%%%%%%%%%%%%%%%%%%%%%%%%%%%%%%%%%%%%%%%%%%%%%%%%%%

%\onecolumngrid

\begin{figure*}[t] 
\includegraphics[width=165mm]{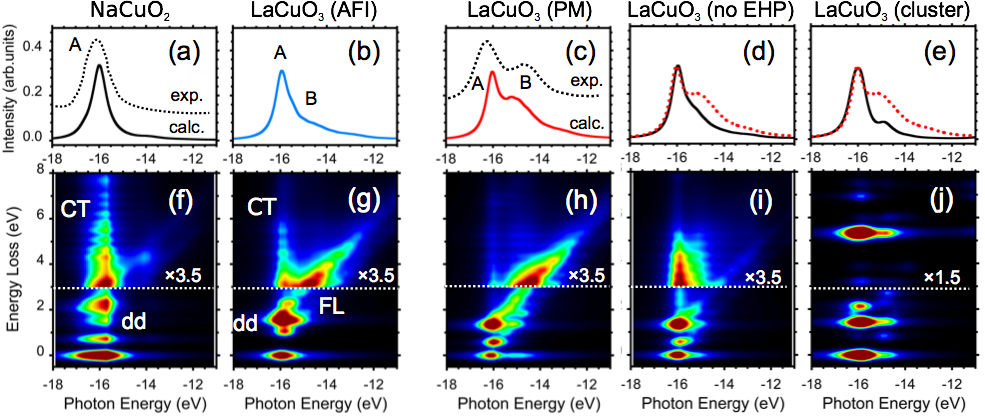}
\caption{The calculated $L_3$-edge XAS and RIXS spectra for (a),(f) NaCuO$_2$ and LaCuO$_3$ in (b),(g) AFI phase, (c),(h) PM phase, (d),(i) no EHP and (e),(j) CuO$_6$ cluster model. The RIXS intensity with $\omega_{\rm loss}\geq~3.0$~eV (horizontal dashed line) are magnified by 3.5 times [1.5 time for (e)]. $\mu_{\rm dc}$=55.64~eV is employed in the calculation.
The LDA+DMFT spectrum for LaCuO$_3$ (PM) is shown by a dotted curve in (d) and (f), for comparison.
%(f) the calculated RIXS by the CuO$_6$ cluster model.
The cluster-model results for NaCuO$_2$ are found in SM~\cite{sm}.
The experimental XAS data (dashed line) for LaCuO$_3$ (PM) and NaCuO$_2$ are taken from Ref.~\cite{Mizokawa98} and Ref.~\cite{Sarma88}. The spectral broadening is considered using a Gaussian of 150~meV for RIXS and a Lorentzian 300~meV for XAS (HWHM).}
\label{fig_rixs}
\end{figure*}

%\twocolumngrid
%%%%%%%%%%%%%%%%%%%%%%%%%%%%%%%%%%%%%%%%%%%%%%%%%%%%%%%%%%%%%%%

In the second step, we compute the RIXS spectra for AIM with DMFT hybridization density and Cu $2p$ core states using the configuration interaction method~\cite{Hariki17}. The hybridization function, which encodes the information of how a given Cu orbital $\gamma$ exchanges electrons with the rest of 
the crystal, can be written as
%%%%%%%%%%%%%%%%%%%%
\begin{equation*}
V_{\gamma}^2(\varepsilon)=\sum_{\alpha}\frac{V^2_{\alpha,\gamma}}{\varepsilon-\varepsilon_{\alpha,\gamma}}.
\end{equation*}
%%%%%%%%%%%%%%%%%%%%
Here, $V_{\alpha,\gamma}$ is the hopping amplitude between the Cu ion and auxiliary orbitals at energies $\varepsilon_{\gamma,\alpha}$~\footnote{In both materials
an independent set of $\alpha$-orbitals can be, to a good approximation, chosen for each Cu orbital.}, 
which represent the effect of the nearest-neighbor oxygen ligands as well as the more  distant atoms~\cite{kotliar06,Hariki17}.
In practice, $V_{\gamma}^2(\varepsilon)$ obtained in the LDA+DMFT calculation is 
%on the Matsubara contour obtained in the DMFT step is
%analytically continued to the real frequency and
represented by 25 discretized bath states $\alpha$ for each Cu orbital $\gamma$~\cite{Hariki17}.
The RIXS intensity at finite temperature $T$ is given
by~\cite{groot_kotani,matsubara05,Kramers25}
%%%%%%%%%%%%%%%%%%%%%%%%%%%%%%%%%%%%%%%%%%%%%%%%
%\begin{eqnarray}
%F_{\rm RIXS}(\omega_{\rm out},\omega_{\rm in})&=&\sum_{n}F^{(n)}_{\rm RIXS}(\omega_{\rm out},\omega_{\rm in})~e^{-E_n/k_{B}T}/Z, \notag 
%\end{eqnarray}
%where 
\begin{equation*}
F_{\rm RIXS}(\omega_{\rm out},\omega_{\rm in})=\sum_{n}F^{(n)}_{\rm RIXS}(\omega_{\rm out},\omega_{\rm in})~e^{-E_n/k_{B}T}/Z,
\end{equation*}
where
\begin{equation}
\label{eq:rixs}
\begin{split}
F^{(n)}_{\rm RIXS}(\omega_{\rm out},\omega_{\rm in})&=\sum_{f} \left| \sum_{m}
     \frac{\langle f | T_{\rm e} | m\rangle \langle m | T_{\rm i} | n \rangle }
     {\omega_{\rm in}+E_n-E_m+i\Gamma_{\rm L}}
      \right|^2 \\
   &\times \delta(\omega_{\rm in}+E_n-\omega_{\rm out}-E_f) \\
   &=\sum_{f} \left| 
     \langle f | T_{\rm e}
     \frac{1}{\omega_{\rm in}+E_n-H_{\rm imp}+i\Gamma_{\rm L}}  T_{\rm i} | n \rangle
      \right|^2 \\
     &\times  \delta(\omega_{\rm in}+E_n-\omega_{\rm out}-E_f)
\end{split}.
\end{equation}
%%%%%%%%%%%%%%%%%%%%%%%%%%%%%%%%%%%%%%%%%%%%%%%%
Here, $|n\rangle$, $|m\rangle$, and $|f\rangle$ are the initial, intermediate and final states with energies 
$E_n$, $E_m$, and $E_f$, respectively, and 
$e^{-E_n/k_BT}/Z$ is the Boltzmann factor with the partition function $Z$. 
%The calculation temperature $T$ is fixed at 290~K.
$\Gamma_{\rm L}$ is the lifetime width of the intermediate state, and 
$T_{\rm i}$ ($T_{\rm e}$) describes the dipole transition for the incident (emitted) photon. 
$H_{\rm imp}$ is the AIM Hamiltonian augmented by the core orbitals and their interaction with Cu 3$d$ orbitals;~see Eq.~(3) in Ref.~\cite{Hariki17}.
In the actual calculation the resolvent formulation on the second line of Eq.~(\ref{eq:rixs}) is used.
We also compute $L$-edge XAS spectra since the intermediate state $|m\rangle$ (corresponding to the final state of XAS) provides
an important clue for interpretation of RIXS spectra.
Details of the calculation can be found in SM~\cite{sm}.

In Figs.~\ref{fig_hyb}(a),\ref{fig_hyb}(b) we show the one-particle (1P) density of states of LaCuO$_3$ and NaCuO$_2$. %calculated by LDA+DMFT.
%\textcolor{blue}{(ORIGINAL) The $\mu_{\text{dc}}$ in the range of 55.64--57.64~eV yields results consistent with earlier valence XPS and Cu $L$-edge XAS studies~\cite{Mizokawa91,Webb91,Mizokawa98,Sarma88}, see also Fig.~\ref{fig_rixs}a,c. The $\mu_{\rm dc}$ dependence of RIXS spectra in this range is rather weak and can be found in SM~\cite{sm}.}
The $\mu_{\text{dc}}$ in the range of 55.64--57.64~eV yields results consistent with Cu $L$-edge XAS studies~\cite{Sarma88,Webb91,Mizokawa98}, see Fig.~\ref{fig_rixs}(a),\ref{fig_rixs}(c).
%The calculated valence spectra of LaCuO$_3$ shows a good agreement with earlier valence study~\cite{Mizokawa98} (Fig.~\ref{fig_hyb}a), while that of NaCuO$_2$ differs noticeably from experiment~\cite{Mizokawa94}~(Fig.~\ref{fig_hyb}b) irrespective of used $\mu_{\text{dc}}$.
The $\mu_{\text{dc}}$ values provide also the best match with earlier valence XPS studies~\cite{Mizokawa98,Mizokawa94} shown in Figs.~\ref{fig_hyb}(a),\ref{fig_hyb}(b).
The deviations from the experimental XPS spectra may be due the uncertainty 
of the relative Cu $3d$ : O $2p$ cross section and the surface sensitivity of XPS.
In NaCuO$_2$, in particular, the surface is prone to contamination leading to 
the appearance of Cu$^{2+}$ ions~\cite{Choudhury15,Sarma88}.
Varying $\mu_{\rm dc}$ within the above range has only a minor impact on the RIXS spectra for both materials and does not affect our conclusions~\cite{sm}.
For LaCuO$_3$, paramagnetic metal (PM) and antiferromagnetic insulator (AFI) solutions can be stabilized, similar to LDA+$U$ studies~\cite{Czyifmmod94}, indicating the Slater nature of the gap. 
Reflecting the unclear experimental situation~\cite{Mizokawa98,Czyifmmod94,Bringley93,Darracq93},
we proceed with both states and use them later to demonstrate the effect of the small gap on RIXS.
NaCuO$_2$ ($E_{\rm gap}\approx 0.5$~eV) has a band-insulator character with a gap present already in the LDA solution~\cite{Singh94,Choudhury15}. Overall, the 1P density of states suggest existing
phase space for continuum of the $p$--$p$ excitations in the few eV range.
The calculated Cu $L_3$-edge XAS and RIXS spectra are shown in Fig.~\ref{fig_rixs}.
The XAS spectrum of NaCuO$_2$ has a single-peak ($A$),
while that of LaCuO$_3$ exhibits an additional shoulder ($B$), enhanced in the PM phase, observed also in experiment for NaCuO$_2$~\cite{Sarma88} and LaCuO$_3$~\cite{Mizokawa98}.
The shoulder $B$ is missing in the calculations on the CuO$_6$ cluster model~\cite{Mizokawa98,Okada99}, see Fig.~\ref{fig_rixs}(e), indicating that the shoulder is not of $d$--$d$ or CT origin but due to a final state delocalized beyond the CuO$_6$ cluster. Such a nonlocal charge excitation is captured by the present approach~\cite{Hariki17}.

%%%%%%%%%%%%%%%%%%%%%%%%%%%%%%%%%%%%%%%%%%
%\begin{figure}
%\begin{center}
%%\includegraphics[width=87mm]{fig_rixs}
% \includegraphics[width=89mm]{fig3_new} 
%\end{center}
%\vspace{-0.4cm}
%\caption{(Color online)
%(a)-(e) the calculated $L$-edge XAS and RIXS for LaCuO$_3$ in PM and AFI phase (left) and NaCuO$_2$ (right).
%The RIXS intensity with $\omega_{\rm loss}\geq~3.0$~eV (white dashed line) are magnified by 3.5 times.
%%(f) the calculated RIXS by the CuO$_6$ cluster model.
%%\textcolor{blue}{The parameter set for LaCuO$_3$ is used in the cluster model. The cluster-model spectrum for NaCuO$_2$ is found in SM~\cite{sm}.}
%The $L_3$-XAS experimental data (dashed line) is taken from Ref.~\cite{Mizokawa98} for LaCuO$_3$ and Ref.~\cite{Sarma88} for NaCuO$_2$. The spectral broadening is considered using a Gaussian of 150~meV for RIXS and a Lorentzian 300~meV for XAS (HWHM).}
%%Magnification of the low energy loss region ($<1.0$) of RIXS map in (f) PM phase and (g) AFI phase of LaCuO$_3$. The FL feature continues to zero energy loss $\omega_{\rm loss}$ in the PM phase.}
%\label{fig_rixs}
%\end{figure}
%%%%%%%%%%%%%%%%%%%%%%%%%%%%%%%%%%%%%%%%%%

The RIXS spectra of NaCuO$_2$ and LaCuO$_3$ are strikingly different.
Tuning $\omega_{\text{in}}$ to the peak A of the XAS, two distinct $d$--$d$ transitions with RL behavior
are found in both compounds, similar to another Cu$^{3+}$ material Zn$_{1-x}$Cu$_x$O~\cite{Thakur10}, followed by CT transitions with higher $\omega_{\text{loss}}$.
However, at higher $\omega_{\text{in}}$ the RIXS of LaCuO$_3$ yields a linear FL feature, with little
difference between the AFI and PM phase. 
The FL feature is suppressed in NaCuO$_2$ resembling the spectrum of the cluster model~\cite{sm}.
The calculated RIXS spectra of LaCuO$_3$ reminds one of the experimental observation on NdNiO$_3$~\cite{Bisogni16} with the FL feature starting at the $\omega_{\text{in}}$ on the $L_3$ main edge and not far above it as in NiO. The continuum of unbound particle-hole pairs in the manner of Ref.~\cite{Bisogni16} explains the FL feature in LaCuO$_3$. 

\begin{figure}
%\vspace{-1.0cm}
\begin{center}
\includegraphics[width=80mm]{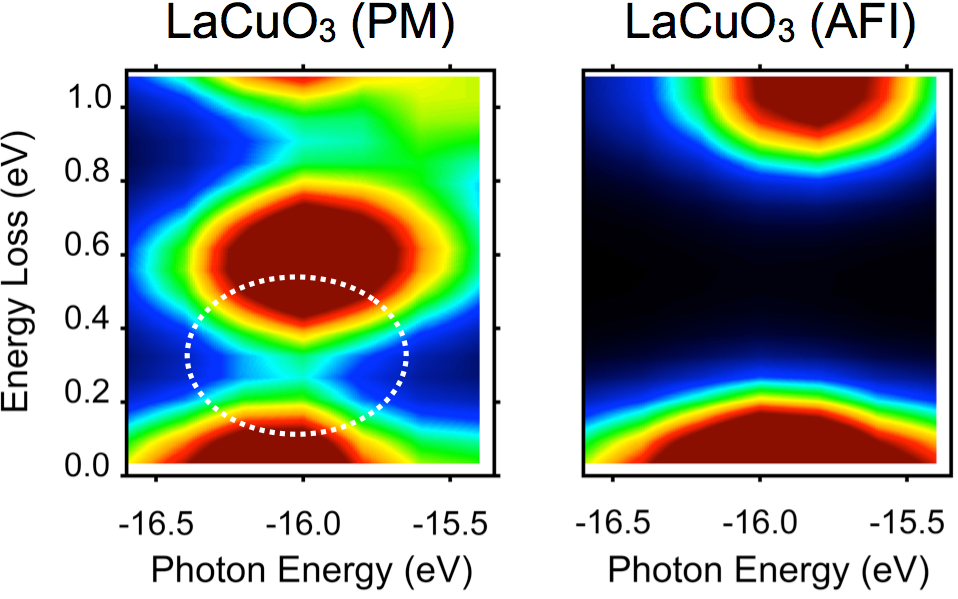}
\end{center}
\vspace{-0.4cm}
\caption{Low $\omega_{\rm loss}$ region ($<1.0$~eV) of the RIXS map in the PM and AFI phases of LaCuO$_3$. The FL feature continues to zero energy loss in the PM phase.}
\label{fig_mag}
\end{figure}

Why is the FL feature missing in NaCuO$_2$ then? The small NaCuO$_2$ gap cannot
explain the absence of visible particle-hole excitations at $\omega_{\rm loss}$ in the 3$-$4~eV range. In fact, the experiment on NdNiO$_3$~\cite{Bisogni16} and the calculations in the PM and AFI phases of LaCuO$_3$ in Fig.~\ref{fig_mag} show that the gap opening affects the FL feature only at low $\omega_{\rm loss}$.
Moreover, the NaCuO$_2$ 1P density of states, Fig.~\ref{fig_hyb}(b), exhibits noticeably higher density of states above and below the gap than LaCuO$_3$, Fig.~\ref{fig_hyb}(a), suggesting a larger phase space for particle-hole excitations.
To answer this question we show, in Figs.~\ref{fig_hyb}(c),\ref{fig_hyb}(d), the hybridization intensities $V_{\gamma}(\varepsilon)$.
%%%%%%%%%%%%%%%
%%%%%%%%%%%%%%%%%%%%%%%%%%%%%%%%%%%%%%%%%%
%%%%%%%%%%%%%%%%%%%%%%%%%%%%%%%%%%%%%%%%%%
%%%%%%%%%%%%%%%%%%
It is instructive to consider the $V_{\gamma}(\varepsilon)$ of the cluster model first~\cite{sm}.
Here, $V_{\gamma}(\varepsilon)$ is a single Dirac $\delta$ function peaked at the energy $\varepsilon_p$ of the ligand orbital, while the O $2p$ density of states exhibits two peaks corresponding to the bonding and antibonding states.

A prominent peak in $V_{\gamma}(\varepsilon)$ (for the $e_g$ orbitals) due to hybridization to nearest-neighbor O ligands is found in both LaCuO$_3$ and NaCuO$_2$. The continuum part of $V_{\gamma}(\varepsilon)$ in the two materials reveals the difference. While in LaCuO$_3$ a substantial hybridization intensity exists in the low-energy region of -2 to 2 eV [blue shadow in Fig.~\ref{fig_hyb}(c)], the $V_{\gamma}(\varepsilon)$ of NaCuO$_2$ resembles that of the cluster model with a weak continuum background. This is how the local quantity $V_{\gamma}(\varepsilon)$, relevant for description of the core-level excitation, encodes the information about bonding and lattice geometry. In LaCuO$_3$, the corner-sharing network of CuO$_6$ octahedra allows electrons and holes to propagate through the strong Cu-O $\sigma$ bonds and thus gives rise to the continuum of $V_{\gamma}(\varepsilon)$. This is not possible in the chains of edge sharing CuO$_4$ plaquettes in NaCuO$_2$, where the neighboring Cu ions form $\sigma$ bonds with orthogonal O $2p$ orbitals, see Fig.~\ref{fig_crys}, and the crystal resembles a collection of weakly coupled CuO$_4$ clusters.

\begin{figure}
%\vspace{-1.0cm}
\begin{center}
  \includegraphics[width=75mm]{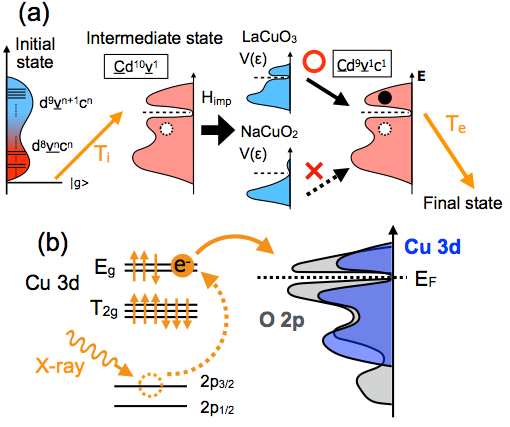}
\end{center}
\vspace{-0.4cm}
\caption{ Schematic of (a) electron-hole pair creation in the"RIXS process of LaCuO$_3$ and NaCuO$_2$, and
(b) CT to the conduction states in the intermediate state.
}
\label{fig_demo}
\end{figure}

How does the hybridization intensity affect the RIXS spectra?
In Eq.~\eqref{eq:rixs}, all intermediate states accessible in the XAS process 
contribute to RIXS in principle. We estimate that the intermediate states with 
$|E_m-E_n-\omega_{\rm in}|\lesssim \Gamma_{\rm L}$, which approximately 
conserve energy in the partial XAS process, dominate while those outside
this range cancel approximately out due to the varying sign of the denominator. 
Such a claim cannot be directly confirmed with the resolvent formulation of Eq.~\eqref{eq:rixs}.
It is, nevertheless, supported by the diagonal shape of the FL feature in the $\omega_{\rm in}$-$\omega_{\rm loss}$ plane, suggesting that a narrow range of intermediate states
are ``excited" that ``decay" into a narrow range of final states with matching electron-hole
excitation.
The small hybridization intensity for $\varepsilon>0$ in NaCuO$_2$ implies that 
(intermediate) states with different numbers of conduction electrons hybridize only weakly with one
another. In LaCuO$_3$ a RIXS process that we schematically write as $d^8+d^9\underline{v} \rightarrow \underline{C}d^{10}\underline{v} + \underline{C}d^9c\underline{v}\rightarrow d^8c\underline{v}$ ends up in a final state that can be characterized as the ground state plus an electron-hole pair in the continuum, see Fig~\ref{fig_demo}(a),
where $\underline{C}$, $\underline{v}$, and $c$ correspond to a hole in 2$p$ core level, in valence bands, and an electron in conduction bands, respectively.
Such processes in NaCuO$_2$ are strongly suppressed since states of the type
$\underline{C}d^{10}\underline{v}$ and $\underline{C}d^9c\underline{v}$ hybridize only weakly.
This is a local expression of the fact that in NaCuO$_2$ a hole transferred from Cu to O has a
small probability to escape the CuO$_4$ cluster.

%%%%%%%%%%%%%%%%%%%%%%%%%%%%%%%%%%%%%%%%%%
%%%%%%%%%%%%%%%%%%%%%%%%%%%%%%%%%%%%%%%%%%

To test this interpretation, we switched off the hybridization to the conduction band in the intermediate states for the PM phase of LaCuO$_3$;~see Fig.~\ref{fig_rixs}(i). In practice, we have set $V_{\gamma}(\varepsilon>0)=0$ in the $H_{\rm imp}$ of Eq.~\eqref{eq:rixs} while keeping $V_{\gamma}(\varepsilon)$ unrestricted in the initial and final states. 
In Fig.~\ref{fig_demo}(b) the hybridization to the conduction band is sketched.
The intensity of the FL feature is dramatically suppressed and the structure of the RIXS spectrum resembles that of NaCuO$_2$.
Although (unchanged) final states with excited electron-hole pairs exist, they cannot be resonantly excited by the RIXS process.
This result supports the interpretation of the FL feature in Ti $L$-edge spectra by Pfaff $et$ $al$.~\cite{Pfaff18} and shows that the nature of hybridization in intermediate states is the dominant factor affecting the intensity of FL feature.
Cutting hybridization to the conduction states affects also the XAS spectrum, Fig.~\ref{fig_rixs}(d),
which loses the shoulder $B$ and overlaps with that of the cluster model, Fig.~\ref{fig_rixs}(e).
This shows that intermediate states with localized and delocalized character coexist in this $\omega_{\rm in}$ region, which leads to coexistence of FL and RL features in the RIXS spectrum.

%The calculation would support the interpretation on the FL feature by Pfaff $et$ $al$.~\cite{Pfaff18} in their Ti $L$-edge RIXS experiments.
%It suggests that XAS is a signature of the appearance of a FL feature and
%that a combined analysis of XAS and RIXS will be useful to study charge excitation in high valence systems.
%It would suggest that the shoulder, lacked in the CT multiplet model, is the signature of 

%%%%%%%%%%%%%%%%%%%%%%%%%%%%%%%%%%%%%%%%%%
%\begin{figure}
%%\vspace{-1.0cm}
%\begin{center}
% %   \includegraphics[width=87mm]{fig_demo.pdf}
%  \includegraphics[width=85mm]{fig4_new}
%\end{center}
%\vspace{-0.4cm}
%\caption{(Color online)
%The calculated $L$-edge (a) XAS and (b) RIXS for LaCuO$_3$ in PM phase
%in which the CT to the conduction states above $E_F$ is removed $only$ in the intermediate state (denoted as ``no EHP").
%(c) Schematic of CT to the conduction states in the intermediate state.
%(d) Schematic of electron-hole pair creation in RIXS process. }
%\label{fig_demo}
%\end{figure}
%%%%%%%%%%%%%%%%%%%%%%%%%%%%%%%%%%%%%%%%%%

In conclusion, we have studied the coexistence of RL and FL features in RIXS spectra of high-valence transition-metal oxides. We have shown how the IAM hybridization function encodes
the information about the lattice environment and that it is the hybridization function 
rather than  one-particle spectral density that determines the presence of  FL feature.
We predict that FL feature is present in the $L$-edge RIXS spectra of LaCuO$_3$ both in the
PM and AFM phases, while it is absent in the isoelectronic NaCuO$_2$, experimental verification of which is highly desirable. We have interpreted this numerical observation in terms 
of crystal geometry.
%We have shown that for geometrical reasons the unbound electron-hole excitations responsible for the 
%FL feature cannot be excited by the RIXS process in NaCuO$_2$. This fact is captured by the hybridization intensity of the Anderson impurity model, which provides the local description 
%of RIXS. 
Comparing the RIXS spectra of PM and AFM LaCuO$_3$, we have shown that the low-$\omega_{\rm loss}$ details of the FL feature are sensitive to opening of a small gap similar to the experimental observation on NdNiO$_3$~\cite{Bisogni16}.
The present results show that the FL component of the RIXS spectra is rather material specific, and its interpretation requires advanced many-body calculations.
%\textcolor{blue}{Further RIXS experimental studies on the FL feature in different lattice geometry are desired, that would open new opportunities to study low-lying electron-hole excitations, especially in high-valence TMO.}

%%%%%%%%%%%%%%%%%%%%%%%%%%%%%%%%%%%%%%%%%%%%%%%%%%%%%%%%%%%%%%%%%%%
\begin{acknowledgments}
The authors thank R. Claessen, M. Sing, P. Scheiderer, T. Uozumi, R.-P. Wang, A. Sotnikov, and J. Fern\'andez Afonso for fruitful discussions.
%A. H, M, W and J. K
The authors are supported by the European Research Council (ERC)
under the European Union's Horizon 2020 research and innovation programme (Grant Agreement No.~646807-EXMAG).
%and T. U is supported by the JSPS KAKENHI Grant Number JP16K05407.
The computational calculations were performed at the Vienna Scientific Cluster (VSC).
\end{acknowledgments}
%%%%%%%%%%%%%%%%%%%%%%%%%%%%%%%%%%%%%%%%%%%%%%%%%%%%%%%%%%%%%%%%%%%

\bibliography{rixs}

\begin{thebibliography}{52}
\bibitem [{cro()}]{cross}%
  \BibitemOpen
  \href@noop {} {\ }\bibinfo {note} {We used the same instrumental broadening
  and the relative Cu $3d$ : O $2p$ cross section as in Ref.~\cite{Mizokawa98},
  where the atomic cross section is taken from
  Ref.~\cite{Scofield76}.}\BibitemShut {Stop}%
\end{thebibliography}

\end{document}

\title{Supplementary material of ``Continuum Charge Excitations in High-Valence Transition-Metal Oxides Revealed by Resonant Inelastic X-ray Scattering''}

\author{Atsushi Hariki}
%\thanks{hariki@ifp.tuwien.ac.at}
\affiliation{Institute for Solid State Physics, TU Wien, 1040 Vienna, Austria}
\author{Mathias Winder}
\affiliation{Institute for Solid State Physics, TU Wien, 1040 Vienna, Austria}
\author{Jan Kune\v{s} }
%\thanks{kunes@ifp.tuwien.ac.at}
\affiliation{Institute for Solid State Physics, TU Wien, 1040 Vienna, Austria}
\affiliation{Institute of Physics,Czech Academy of Sciences, Na Slovance 2, 182 21 Praha 8, Czechia}

%\date{\today}%a

\maketitle
\renewcommand{\theequation}{S.\arabic{equation}}
\renewcommand{\thefigure}{S.\arabic{figure}}
\renewcommand{\thetable}{S.\arabic{table}}
\renewcommand{\bibnumfmt}[1]{[s#1]}
\renewcommand{\citenumfont}[1]{s#1}

%\clearpage
%\newpage
\subsection{Computation of RIXS intensities}
\label{sec_rixs}

$L$-edge RIXS cross-section $\frac{\delta^2\sigma}{\delta\Omega\delta \omega} \propto F_{\rm RIXS}$
is described by the Kramers-Heisenberg formula~\cite{groot_kotani,Kramers25},
%%%%%%%%%%%%%%%%%%%%%%%%%%%%%%%%%%%%%%%%%%%%%%%%%
\begin{equation*}
F_{\rm RIXS}(\omega_{\rm out},\omega_{\rm in})=\sum_{n}F^{(n)}_{\rm RIXS}(\omega_{\rm out},\omega_{\rm in})~e^{-E_n/k_{B}T}/Z,
\end{equation*}
%%%%%%%%%%%%%%%%%%%%%%%%%%%%%%%%%%%%%%%%%%%%%%%%%
where
%%%%%%%%%%%%%%%%%%%%%%%%%%%%%%%%%%%%%%%%%%%%%%%%%
\begin{eqnarray}
	F^{(n)}_{\rm RIXS}(\omega_{\rm in},\omega_{\rm out}) 
	&=&
	\sum_{f} \Big{|} \sum_m \frac{ \langle f | \hat{T}_{e} |m\rangle \langle m | \hat{T}_{i}| n \rangle}{\omega_{\rm in}+E_n-E_m+i\Gamma_{\rm L}} \Big{|} ^{2}
	\delta(\omega_{\rm in}+E_n-\omega_{\rm out}-E_f) \nonumber \\
	\vspace{+0.5cm}
    &=&
	\sum_{f} \Big{|}  \langle f | \hat{T}_{e}  \frac{ 1 }{\omega_{\rm in}+E_n-\hat{H}+i\Gamma_{\rm L}} \hat{T}_{i} | n \rangle \Big{|} ^{2}
	\delta(\omega_{\rm in}+E_n-\omega_{\rm out}-E_f). \nonumber 
\end{eqnarray}
%%%%%%%%%%%%%%%%%%%%%%%%%%%%%%%%%%%%%%%%%%%%%%%%%
Here, $|n\rangle$, $|m\rangle$, and $|f\rangle$ represent
the initial, intermediate, and final states with energies $E_n$, $E_m$, and $E_f$, respectively.
$e^{-E_n/k_BT}/Z$ is the Boltzmann factor with the partition function $Z$.
$\Gamma_{\rm L}$ is the lifetime width, and 
$\hat{T}_i$ ($\hat{T}_e$) is the transition operator that describes the x-ray absorption (emission) in the second-order optical process.
In actual calculations, $\Gamma_{\rm L}$ is replaced by a constant value of 0.3~eV.
The geometry setting including the x-ray polarization is encoded in the operators $\hat{T}_i$ and $\hat{T}_e$.
In the present study, we employed the polarization configuration, where the polarization vector of x-rays is perpendicular to the scattering plane~\cite{matsubara05}.
%The scattering plane is assumed be in $xz$ plane for LaCuO$_3$ and the 45$^\circ$-rotated plane from that (along $z$ axis) for NaCuO$_2$.
The incident (emitted) x-ray has the energy $\omega_{\rm in}$  ($\omega_{\rm out}$),
and energy loss is given by $\omega_{\rm loss}=\omega_{\rm in}-\omega_{\rm out}$.
The operator $\hat{H}$ is the Hamiltonian of the whole system and, in the LDA+DMFT approach, 
$\hat{H}$ is replaced by that of the Anderson impurity model (AIM) with the DMFT hybridization intensity~\cite{Hariki17}.
The AIM Hamiltonian $\hat{H}_{\rm imp}$ is augmented by the 2$p$ core orbitals and their interaction with Cu 3$d$ orbitals, 
and thus has the form
%%%%%%%%%%%%%%%%%%%%%%%%%%%%%%%%%%%%%%%%%%%%%%%%%
\begin{equation*}
\hat{H}_{\rm imp}= \hat{H}_{\rm TM}+ \hat{H}_{\rm hyb}.
\end{equation*}
%%%%%%%%%%%%%%%%%%%%%%%%%%%%%%%%%%%%%%%%%%%%%%%%%
The on-site Hamiltonian $\hat{H}_{\rm TM}$ is given as
%%%%%%%%%%%%%%%%%%%%%%%%%%%%%%%%%%%%%%%%%%%%%%%%%%%%%%%%%%%%%
\begin{align}
\hat{H}_{\rm TM}
&=\sum_{\gamma,\sigma}\tilde{\varepsilon}_{d} (\gamma) \hat{d}_{\gamma\sigma}^{\, \dagger} \hat{d}_{\gamma\sigma} %\notag \\
 + U_{dd} \sum_{\gamma\sigma > \gamma'\sigma'}
  \hat{d}_{\gamma\sigma}^{\, \dagger} \hat{d}_{\gamma\sigma} \hat{d}_{\gamma'\sigma'}^{\, \dagger} \hat{d}_{\gamma'\sigma'}  %\\
 - U_{dc} \sum_{\gamma,\sigma , \,\zeta,\eta} 
  \hat{d}_{\gamma\sigma}^{\, \dagger} \hat{d}_{\gamma\sigma} (1-\hat{c}_{\zeta\eta}^{\, \dagger} \hat{c}_{\zeta\eta}) %\notag %\\
 %+ H_{\rm S-O} + H_{\rm Coulomb},
 +\hat{H}_{\rm multiplet}, \notag
\label{eq:Imp_Hamiltonian}
\end{align}
%%%%%%%%%%%%%%%%%%%%%%%%%%%%%%%%%%%%%%%%%%%%%%%%%%%%%%%%%%%%%
Here, ${\hat{d}_{\gamma\sigma}^{\, \dagger}}$ (${\hat{d}_{\gamma\sigma}}$) and ${\hat{c}_{\zeta\eta}^{\, \dagger}}$ (${\hat{c}_{\zeta\eta}}$) are the electron creation (annihilation) operators  for Cu 3$d$ and 2$p$ electrons, respectively.
The ${\gamma}$ (${\zeta}$) and ${\sigma}$ (${\eta}$) are the Cu ${3d}$ (2$p$) orbital and  spin indices.
The Cu $3d$ site energies $\tilde\varepsilon_{d}(\gamma)=\varepsilon_{d}(\gamma)-\mu_{\rm dc}$ are the energies
of the Wannier states $\varepsilon_{d}(\gamma)$ shifted by the double-counting correction $\mu_{\rm dc}$. 
The isotropic part of the 3$d$-3$d$ ($U_{dd}$) and 2$p$-3$d$ ($U_{dc}$) interactions are shown explicitly,
while terms containing higher Slater integrals and the spin-orbit (SO) interaction are contained in $\hat{H}_{\rm multiplet}$.
In the present study $U_{dc}$=8.5~eV is employed.
The SO coupling within the 2$p$ shell and the anisotropic part of the 2$p$-3$d$ interaction parameters $F^k$, $G^k$ are calculated with an atomic Hartree-Fock code. 
The $F^k$ and $G^k$ values are scaled down to 85\%
of their actual values to simulate the effect of intra atomic-configuration interaction 
from higher basis configurations neglected in the atomic calculation,
which is a successful empirical treatment \cite{groot_kotani,cowan,Sugar72,matsubara05}.
The $\hat{H}_{\rm hyb}$ describes hybridization with the fermionic bath and is given as 
%%%%%%%%%%%%%%%%%%%%%%%%%%%%%%%%%%%%%%%%%%%%%%%%%%%%%%%%%%%%%
\begin{equation}
\hat{H}_{\rm hyb} = \sum_{\alpha,\gamma,\sigma} \epsilon_{\alpha,\gamma,\sigma} \hat{v}_{\alpha\gamma\sigma}^\dagger \hat{v}_{\alpha\gamma\sigma} + \sum_{\alpha,\gamma,\sigma} V_{\alpha,\gamma,\sigma}(\hat{d}_{\gamma\sigma}^\dagger \hat{v}_{\alpha\gamma\sigma} + \hat{v}_{\alpha\gamma\sigma}^\dagger \hat{d}_{\gamma\sigma} )\notag
\end{equation}
%%%%%%%%%%%%%%%%%%%%%%%%%%%%%%%%%%%%%%%%%%%%%%%%%%%%%%%%%%%%%
The first term represents the energies of the auxiliary orbitals and the second term 
describes the hopping between the Cu 3$d$ ion and the auxiliary orbitals with the amplitude $V_{\alpha,\gamma,\sigma}$.
Here ${\hat{v}_{\alpha\gamma\sigma}^{\, \dagger}}$ (${\hat{v}_{\alpha\gamma\sigma}}$) is the creation (annihilation) operator for the auxiliary state with energy $\epsilon_{\alpha,\gamma,\sigma}$.
The hybridization is assumed to be orbital (and spin) diagonal, which is a good approximation in the studied compounds.
The configuration interaction scheme with 25 discretized-bath states representing the DMFT hybridization density
is employed to compute RIXS intensities.
In the discretization, we have employed uniform energy meshes
in the intervals with sizable hybridization intensity (typically defined as larger than 5\% of the peak value)~\cite{Hariki17}.
The initial state is computed using the Lanczos method. 
The intermediate state propagated by the resolvent as $(\omega_{\rm in}+E_n-\hat{H}_{\rm imp}+i\Gamma_{\rm L})^{-1}\hat{T}_{i}|n\rangle$ is computed using the conjugate-gradient method,
and then its overlap to the final states are computed again using the Lanczos method.

\begin{figure}[ht]
    \includegraphics[width=0.70\columnwidth]{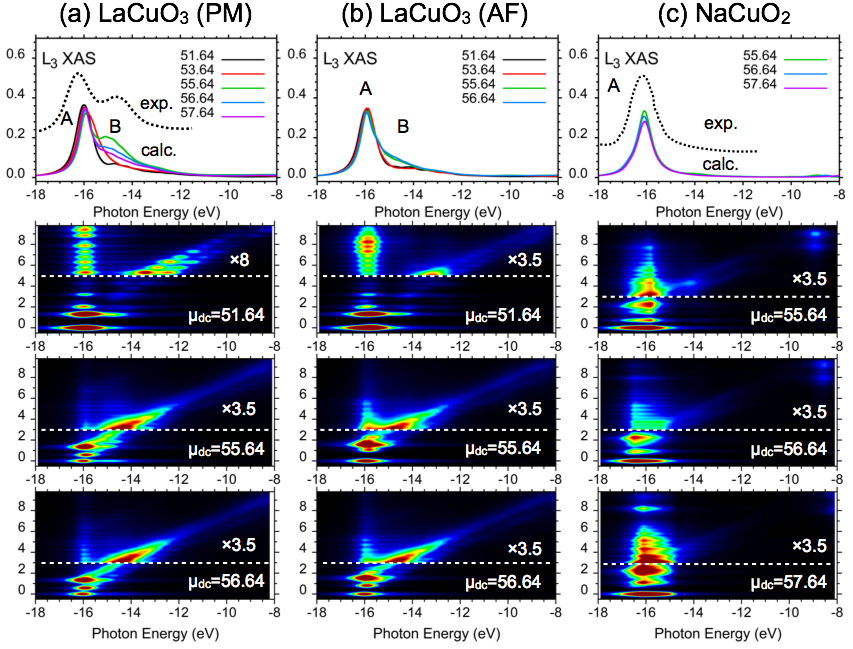}
	\caption{Calculated $L_3$-edge XAS and RIXS for (a) LaCuO$_3$ (PM), (b) LaCuO$_3$ (AF), and (c) NaCuO$_2$. The experimental XAS data (dashed line) for LaCuO$_3$ (PM) and NaCuO$_2$ are taken from Ref.~\cite{Mizokawa98} and Ref.~\cite{Sarma88}. 
	}\label{figS1}
\end{figure}

\subsection{Double-counting correction dependence of XAS, RIXS and valance spectra}
\begin{figure}[ht]
    \includegraphics[width=0.90\columnwidth]{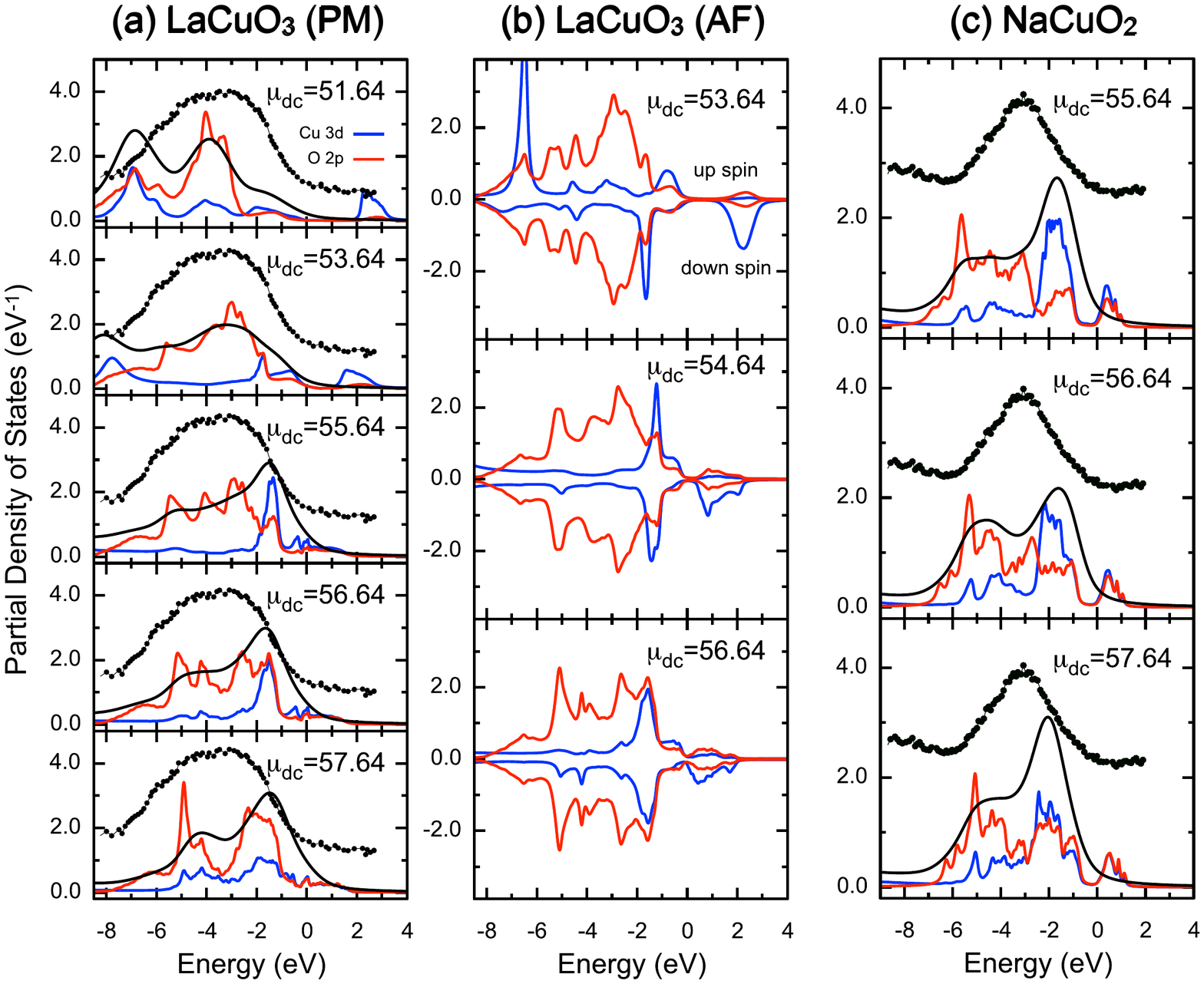}
	\caption{
    Calculated 1P density of states of (d) LaCuO$_3$ (PM), (e) LaCuO$_3$ (AF), and (f) NaCuO$_2$ for different double-counting corrections $\mu_{\rm dc}$. The calculated valence spectra (black line) are shown together. The experimental XPS data of LaCuO$_3$ and NaCuO$_2$ are taken from Ref~\cite{Mizokawa98} and \cite{Mizokawa94}, respectively.
	}\label{figS2}
\end{figure}

Figure~\ref{figS1} shows Cu $L$-edge XAS and RIXS calculated for different double-counting corrections $\mu_{\rm dc}$ in LaCuO$_3$ and NaCuO$_2$. 
The double-counting corrects for the $d$--$d$ interaction present in the LDA calculation.
As one could find in the $dp$ Hamiltonian in the main text, $\mu_{\rm dc}$ renormalizes the $p$--$d$ splitting and a larger $\mu_{\rm dc}$ reduces the splitting, i.e.~the charge-transfer (CT) energy.
The XAS spectra of LaCuO$_3$ with $\mu_{\text{dc}}=51.64$~eV, deep in the Mott insulating regime, shows a sharp main line (A) and a weak multiplet feature around $-14\sim-15$~eV, resembling Ni $L$-edge XAS of NiO ($3d^8$ electronic configuration)~\cite{Alders98}.     
With $\mu_{\rm dc}$ increase (i.e.,~CT energy decrease), a new shoulder feature $B$ develops in both antiferromagnetic (AF) and paramagnetic (PM) phases. The metal-insulator transition occurs at $\mu_{\text{dc}}\sim 55.64$~eV in the PM solution, while a gap survives in the AF solution for $\mu_{\rm dc}\lesssim 56.74$~eV.  
The $\mu_{\text{dc}}$=55.64--57.64~eV yields a reasonable agreement to the experimental XAS data measured using
the total electron yield method~\cite{Mizokawa98}. The XAS of NaCuO$_2$ shows a rather weak $\mu_{dc}$ dependence and does not show the shoulder $B$, consistent with the experimental observation~\cite{Sarma88}. The estimated $\mu_{\rm dc}$ values give similar RIXS features in LaCuO$_3$, showing coexisting fluorescencelike (FL) and Ramman-like features across the Cu $L_3$ main line,
while a small $\mu_{\rm dc}$, see $\mu_{\rm dc}=51.64$~eV, gives a FL feature that appears far above the $L_3$ main line as in NiO~\cite{Ghiringhelli05}.
The calculated valance XPS with the estimated $\mu_{\text{dc}}$ give a fair agreement to early photoemission data~\cite{Mizokawa94,Mizokawa98}, see Fig.~\ref{figS2}.

\subsection{Hybridization density $V_{\gamma}^2(\varepsilon)$, XAS and RIXS spectra in cluster model}

\begin{figure}[ht]
    \vspace{0.5cm}
    \includegraphics[width=0.4\columnwidth]{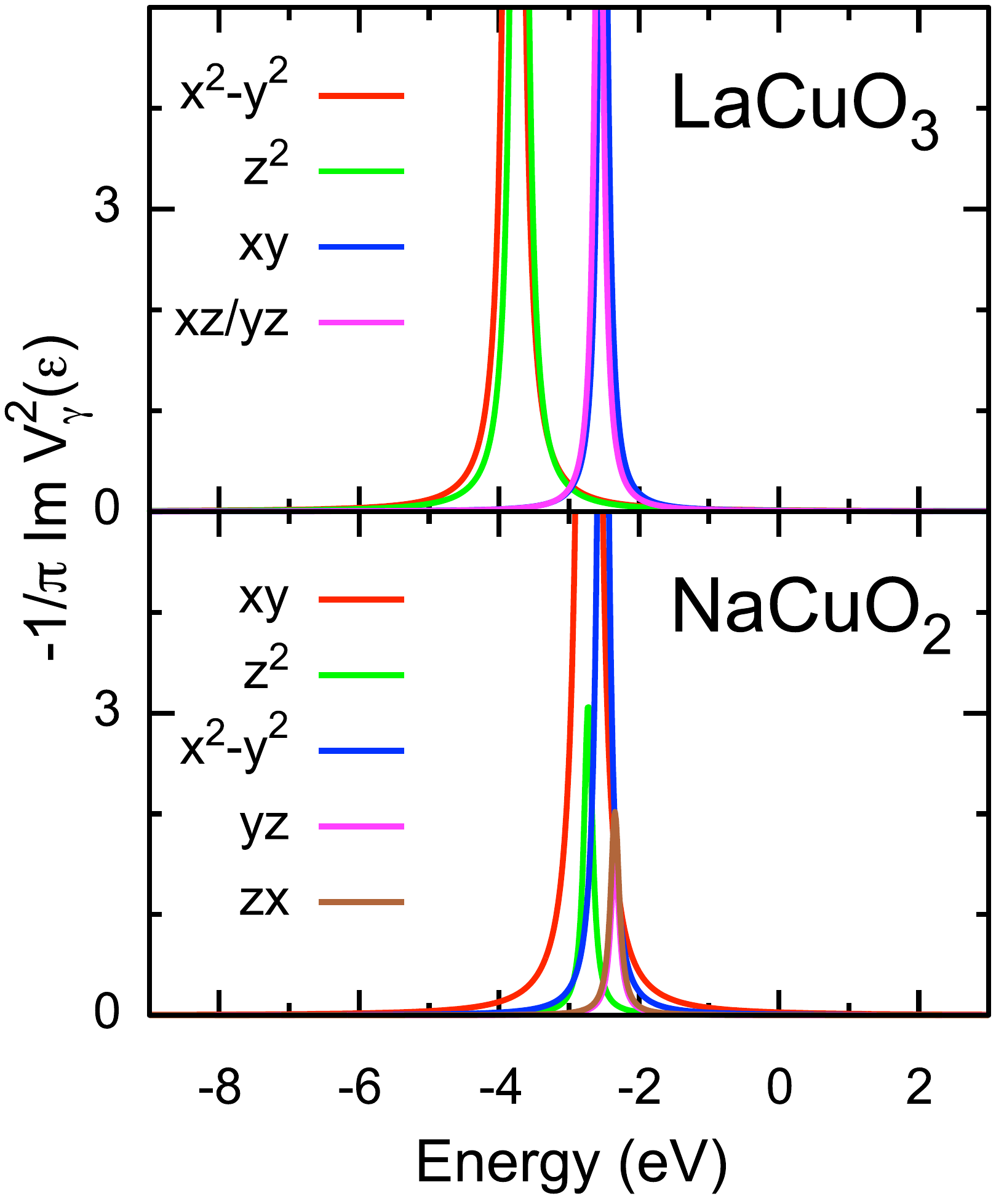}
	\caption{The hybridization function  of LaCuO$_3$ and NaCuO$_2$ in the cluster model.
	}\label{figS3}
\end{figure}

Figure~\ref{figS3} shows the hybridization density $V_{\gamma}^2(\varepsilon)$ of the studied compounds in the cluster model. Considering the crystal structure, see Fig.~1 in the manuscript, we here assumed CuO$_6$ and CuO$_4$ cluster for LaCuO$_3$ and NaCuO$_2$, respectively. The $V_{\gamma}^2(\varepsilon)$ shows a single-peak feature, as expected, while the energy position of the peaks depends on $\gamma$ because of the crystal-field splitting in O 2$p$ states. 

\begin{figure}[ht]
    \includegraphics[width=0.8\columnwidth]{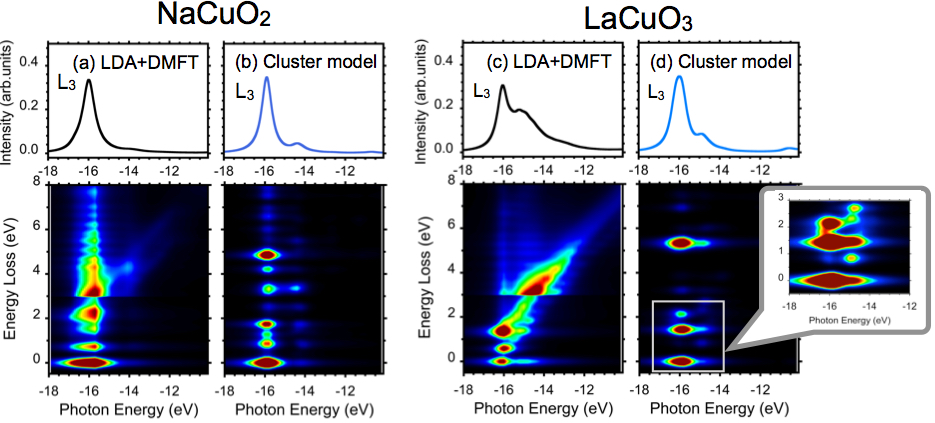}
	\caption{The $L_3$ XAS and RIXS spectra for NaCuO$_2$ and LaCuO$_3$ calculated by the LDA+DMFT approach and the cluster model. The magnification of the RL features in the cluster-model spectrum is shown, for comparison.
	}\label{figS4}
\end{figure}

Figure~\ref{figS4} shows the $L_3$ XAS and RIXS spectra computed by the LDA+DMFT approach and the cluster model. For comparison, a magnified RIXS map of the RL features is shown for LaCuO$_3$.  We find a reasonable agreement in the energy positions of the RL features ($\lesssim$~2.0 eV) between the two methods. However, the intensities and the resonance photon energies of the RL features differ between the two methods. These depend, in general, on the character of the intermediate states of the RIXS process. We remind that the cluster model does not describe the intermediate states of LaCuO$_3$ properly, which leads to the absence of the shoulder~B in the XAS spectrum of the cluster model (as compared to experiment). 

%\clearpage
\bibliography{sm}